\documentclass[showpacs,prb,preprint]{revtex4-1}
\usepackage{graphicx}
\usepackage{amsmath}
\usepackage{color}
\begin{document}

\textbf{{Accepted for publications in J. Appl. Phys.(2012)}}

\title{A first-principles investigation on the effects of magnetism  
on the Bain transformation of $\alpha$-phase FeNi systems}
\author{Gul Rahman}
\altaffiliation[Current Address: ]{Department of Physics, Quaid-i-Azim University,
Isamabad, Pakistan}
\author{In Gee Kim}
\email[Corresponding author: ]{igkim@postech.ac.kr}
\affiliation{Graduate Institute of Ferrous Technology, 
Pohang University of Science and Technology, 
Pohang 790-784, Republic of Korea}
\author{H. K. D. H. Bhadeshia}
\affiliation{Graduate Institute of Ferrous Technology, 
Pohang University of Science and Technology,
Pohang 790-784, Republic of Korea}
\affiliation{Department of Materials Science and Metallurgy, 
University of Cambridge,Cambridge CB2 3QZ, U.K.}


\begin{abstract}
The effects of magnetism on the Bain transformation of $\alpha$-phase FeNi systems are investigated 
by using the full potential linearized augmented plane wave (FLAPW) method based on 
the generalized gradient approximation (GGA).
We found that Ni impurity in bcc Fe increases the lattice constant in ferromagnetic (FM) states,
but not in the nonmagnetic (NM) states.
The shear modulus $G$ and Young's modulus $E$ of bcc Fe are also increased by raising the concentration of nickel. 
All the compositions considered show high  shear anisotropy and the  ratio of the bulk to shear modulus is greater than $1.75$ implying ductility. 
The mean sound velocities in the 
$\left[ 100 \right]$ directions are greater than in the $\left[ 110 \right]$ directions. 
The Bain transformation, which is a component of martensitic transformation, has also been studied to reveal that Ni$_{x}$Fe$_{1-x}$ alloys are elastically unstable in the NM states, 
but not so in the FM states. 
The electronic structures explain these results in terms of   
the density of states at the Fermi level. 
It is evident that magnetism cannot be neglected when dealing with 
the Bain transformation in iron and its alloys.
\end{abstract}

\maketitle

\section{Introduction}
Nickel and iron form substantial solid solutions over the complete range of compositions. 
The addition of nickel to iron enhances the strength and toughness of iron. 
In its $\alpha$-phase, the iron-nickel alloy forms the base of structural steel; henceforth, 
the mechanical properties of these alloys received more attention as compared to other properties.
The body-centered cubic (bcc) $\alpha$-phase is stable from pure iron to $\sim$ 10 \,\% Ni and  
increasing the Ni concentration in $\alpha$-Fe reduces its stability with respect to 
$\gamma$-Fe which is why nickel is designated a $\gamma$-stabilizer.\cite{owen}
Fe-Ni alloys are also of interest in connection with Invar effect and martensitic transformation
at low temperatures.\cite{williams,meyer}
Fe-Ni alloys are amongst the most studied magnetic materials and are important 
in understanding the mechanical and magnetic properties of steels. 
Considerable efforts have been made to understand the stability of magnetic materials in terms of 
magnetism and Bain transformation,\cite{Olson,Ekman,Friak,Hsueh,Okatov} 
but with the focus on  Bain path of either pure bcc Fe, or fcc Fe, or hcp Fe, 
rather than on the consequences of the presence of nickel as a solute.

Much attention has also been given in the past to Fe-Ni alloys near Invar compositions or 
ordered compounds (Fe${_3}$Ni, FeNi, or FeNi${_3}$), 
but not dilute alloys with less than 10\,\% nickel.\cite{williams,meyer,filho,mishin}
Recently,~\cite{RahmanPRB2010} we studied the electronic and magnetic structures of transition metal impurities in bcc Fe where we also found that Ni impurity enhances the magnetism of bcc Fe.
To further elucidate the effect of Ni on iron, we examine here the elastic properties, thermodynamics, 
and the effects of magnetism on the Bain path using first-principles calculations. 

\section{Computation Model and Methods}
\subsection{Computation Model}
We considered a $3\times 3\times 3$ supercell of the primitive cell of bcc Fe, 
which contains  27 atomic sites. 
The centered Fe atom is replaced by the Ni atom; we denote the model as Ni$_{1}$Fe$_{26}$.
If we replace the central Fe atom by Ni, then the space group is 
$Im\overline{3}m$ (space group $\#229$), 
and the Ni atom is at the ($2a$) Wyckoff position, 
its 8 nearest neighbor Fe atoms (denoted as Fe1) are on the ($16f$) Wyckoff sites, 
the 6 second nearest neighbors (denoted as Fe2)  are on the ($12e$) Wyckoff sites, 
and the 12 third nearest neighbors (denoted as Fe3) are on the ($24h$) Wyckoff sites.
A $2\times 2\times 2$ supercell of a conventional cubic cell of bcc Fe, 
which contains 16 atoms, was also considered and this system is denoted as Ni$_{1}$Fe$_{15}$. 
The supercell is a simple cubic unit cell, with lattice constant twice that of the bcc structure. 
The space group is $Pm\overline{3}m$ (space group $\# 221$), and the Ni atom is on the ($1a$) Wyckoff position. 
The neighboring atoms are on ($8g$), ($3c$), ($3d$), and ($1b$) sites,
in order of increasing distance from the central Ni atom. 
The corresponding Ni concentrations are  $3.7$ and $6.25$\,at.\%  for $3\times 3\times 3$  
and $2\times 2\times 2$ supercells, respectively. 
Pure bcc Fe, bcc Ni, and fcc Ni as references were also considered. 

\subsection{Electronic structure calculation method}
The Kohn-Sham equation was solved self-consistently in terms of 
the total energy all-electron full-potential linearized augmented plane-wave (FLAPW) method\cite{flapw,flapw1} 
based on the generalized gradient approximation (GGA) for the exchange-correlation potential.\cite{GGA} 
The integrations over the three dimensional Brillouin zone (3D-BZ) were performed by
the improved tetrahedron method~\cite{jhlee}
 over a 13$\times$13$\times$13 Monkhorst-Pack mesh~\cite{monk} in the 3D-BZ, 
which corresponds to 84 \textbf{k}-points inside the irreducible wedge of the 3D-BZ. 
The linearized augmented plane-wave (LAPW) basis set was expanded using 
a plane wave with an energy cutoff at 4 ($2\pi/a$), where $a$ is the lattice parameter. 
Lattice harmonics with $l \leq 8$  were employed to expand the charge density, potential, 
and wave functions inside each muffin-tin (MT) sphere with the radius of 2.20 a.u. 
for both Fe and Ni atoms. 
The star-function cutoff of $16.73$ ($2\pi/a$) was employed for depicting the charge density 
and potential in the interstitial region.
The core electrons were treated fully relativistically, 
and the valence electrons were treated scalar relativistically. 
To ensure the orthogonality between core and valence states,
we employed the explicit orthogonalization (XO) method.\cite{XO}
Self-consistency was assumed when the difference between 
input and output charge (spin) density is less than 
$1.0\times10^{-5}$\,electrons/a.u.$^{3}$ 
The convergence of these computational parameters was carefully checked.\cite{Seo}

\subsection{Methods for elastic properties}
Calculations were carried out on nonmagnetic (NM) and ferromagnetic (FM) states,
to find the optimized lattice constant by fitting the total energy data at various lattice constants
to the Birch-Murnaghan equation of state (EOS),\cite{birch}
which also provides the bulk modulus $B$. 

Since the considered systems are cubic, there are only three independent elastic constants; 
$C_{11}$, $C_{12}$, and $C_{44}$. 
These elastic constants can be determined by calculating the total energy as a function of the shears described below. 
To determine $C_{11}$ and $C_{12}$, 
we considered a volume-conserved tetragonal strain in such a way as
to modify the cubic crystal axes by applying the following strain matrix,
\begin{displaymath}
        \left( \begin{array}{ccc}
                     1+\delta & 0 & 0 \\
                     0 & 1+\delta & 0 \\
                     0 & 0 & (1+\delta)^{-2} \\
                    \end{array}
             \right)
\end{displaymath}
where $\delta$ is the amount of strain $\pm 0.03$ imposed on the crystal. The change in the strain energy density ($u$) 
as a function of strain is given by
\begin{equation}
\label{cprime}
 u=6C^{\prime}\delta^{2}+O(\delta^{3}),
\end{equation}
where $C^{\prime}$ is the tetragonal shear constant 
defined by ($C_{11}$-$C_{12}$)/2. 
By calculating $C^{\prime}$ and bulk modulus relation 
$B$ =$\frac{1}{3}$($C_{11}$+$2C_{12}$) one can estimate $C_{11}$ and $C_{12}$. 
We must keep in mind that the bulk modulus $B$ was calculated by using the EOS.\cite{birch}
To determine $C_{44}$, the following volume-conserved orthorhombic distortion 
\begin{displaymath}
        \left( \begin{array}{cccccc}
                     1   &      & & \delta  & &0  \\
                     \delta  & && 1         & &0 \\
                      0     &   &&  0         && (1-\delta^{2})^{-1}
                    \end{array}
             \right)
\end{displaymath}
was considered and $C_{44}$ can be calculated from 
\begin{equation}
\label{cprime2}
u=2C_{44}\delta^{2}+O(\delta^{4}) .
\end{equation}
The above elastic constants can be put into a more general way, \textit{i.e}, 
the single-crystal shear moduli for the $\left\{100\right\}$ plane along the $\left[010\right]$ direction 
and for the $\left\{110\right\}$ plane along the $\left[1\overline{1}0\right]$ direction are 
simply given by $G_{\left\{100\right\}}=C_{44}$ 
and $G_{\left\{110\right\}}=(C_{11}-C_{12})/2$, 
respectively.\cite{Wohlfarth}
Once  $C_{12}$ and $C_{44}$ are known, it becomes possible to calculate the Cauchy pressure $C_{P}=C_{12}-C_{44}$.

The shear modulus $G$, Young's modulus $E$, Poisson's ratio $\nu$, 
and shear anisotropy factor $A$ for polycrystalline aggregates can be calculated 
from the elastic constants. 
The shear modulus is bounded by the Reuss' $G_\mathrm{R}$ modulus and Voigt's $G_\mathrm{V}$ one
which represent their lower and upper limits, respectively.\cite{voigt,reuss}
For cubic lattices these moduli are determined by
\begin{equation}
G_{R}=\frac{5(C_{11}-C_{12})C_{44}}{4C_{44}+3(C_{11}-C_{12})}  , 
\end{equation}
\noindent   and 
\begin{equation}
  G_{V}= {\frac{(C_{11}-C_{12}+3C_{44})}{5}} .
\end{equation}
Despite wide usage, neither Russ' nor Voigt's relation is believed to be exact. 
Hill\cite{hill} suggested an averaging by arithmetic mean of  $G_\mathrm{R}$ and $G_\mathrm{V}$
\begin{equation}
\label{shear}
 G= \frac{1}{2}(G_{R}+G_{V}) .
\end{equation}

The bulk and shear moduli are used to calculate the Young's modulus 
\begin{equation}
\label{Young}
E=\frac{9BG}{3B+G},
\end{equation}
and the Poisson's ratio 
\begin{equation}
\label{Poisson}
\nu=\frac{3B-E}{6B}.
\end{equation}
The shear anisotropic factor $A$ can be calculated by 
using the following relation\cite{zener}
\begin{equation}
\label{anisotropy}
A=\frac{2C_{44}}{C_{11}-C_{12}}.
\end{equation}

\subsection{Methods for thermodynamic data}
Using the equilibrium lattice constant, the formation enthalpy $\Delta H$ per atom of 
Ni$_{n}$Fe$_{m}$ was calculated as follows:
\begin{equation}
\Delta H=
\frac{ H\left( \mathrm{Ni}_{n}\mathrm{Fe}_{m} \right)
- m H \left( \mathrm {Fe} \right)
- n H \left( \mathrm{Ni} \right)}{m+n},
\label{formationenthalpy}
\end{equation}
where $H(\mathrm{Ni}_{n}\mathrm{Fe}_{m})$ is the enthalpy of 
Ni$_{n}$Fe$_{m}$ with $m=26, 15$ and $n=1$, and 
$H(\mathrm{Fe})$ and $H(\mathrm{Ni})$ are the total energies per atom of 
the ground states of bcc Fe and fcc Ni, respectively.

The Debye model\cite{Debye} was used to calculate 
the mean isotropic velocity $v_\mathrm{m}$  which is given by
\begin{equation}
\label{vmeq}
v_{m}=\left[\frac{1}{3}\left( \frac{1}{v^{3}_\bot}+\frac{2}{v^{3}_\|}\right)\right]^{-1/3} ,
\end{equation}
where the longitudinal $v_\bot$ and transverse $v_\|$ velocities are given by     
\begin{equation}
\label{vmeq1}
v_\bot=\sqrt{(B+4G/3)/\rho} ,   
\end{equation}
and
\begin{equation}
\label{vmeq2}
v_\|=\sqrt{G/\rho} .
\end{equation}
where $\rho$ denotes the density of material. 

Once we know all the elastic constants of a cubic system,
we can also examine the behavior of sound velocities in different crystallographic directions, 
\textit{e.g.}, [100], [110], or [111] directions.\cite{Kittel} 
The velocity of longitudinal ($v_{l}$) and transverse ($v_{t}$) elastic wave
in the [100] direction is given by
\begin{equation}
\label{sound1}
v_{l}=\left(\frac{C_{11}}{\rho}\right)^{1/2},
\;\; v_{t}=\left(\frac{C_{44}}{\rho}\right)^{1/2},
\end{equation}
Similarly, for the [110] direction
\begin{equation}
\label{sound2}
v_{l}=\left(\frac{C_{11}+C_{12}+2C_{44}}{2\rho}\right)^{1/2},
\;\; v_{t}=\left(\frac{C_{11}-C_{12}}{2\rho}\right)^{1/2},
\end{equation}
and for the [111] direction, the velocities can be expressed as
\begin{equation}
\label{sound3}
v_{l}=\left(\frac{C_{11}+2C_{12}+4C_{44}}{3\rho}\right)^{1/2},
\;\; v_{t}=\left(\frac{C_{11}-C_{12}+C_{44}}{3\rho}\right)^{1/2} .
\end{equation}
$v_{l}$ and $v_{t}$ were used to distinguish it from the isotropic velocities given
in Eqs.~(\ref{vmeq}--\ref{vmeq2}).

\section{Results and Discussion}
\subsection{Formation Enthalpy}
\begin{table}[ht]
\caption{The calculated lattice constant ($a$) in  units of \AA$\,$ of bcc Fe, bcc Ni, fcc Ni, 
and bcc Ni$_{1}$Fe$_{26}$ and Ni$_{1}$Fe$_{15}$. 
Both the nonmagnetic (NM) and ferromagnetic (FM) results are listed. 
The experimental values (Expt.) of the bcc Fe,\cite{monk}
bcc Ni,\cite{bccNi} fcc Ni,\cite{monk}
 and bcc Ni$_{1}$Fe$_{26}$,\cite{mckeehan} 
and the other theoretical results\cite{guo} (Other) are also listed for comparison.} 
\begin{center} 
\begin{tabular}{cccccccccccc}
\hline \hline
&$\;\;\;$bcc Fe & $\;\;\;$bcc Ni & $\;\;\;$fcc Ni & $ \;\;\;$Ni$_{1}$Fe$_{26}$ &  $\;\;\;$Ni$_{1}$Fe$_{15}$ \\ 
\hline
NM &    2.76  &     2.79 &       3.51  &     2.76 &    2.76    \\
FM &    2.83   & 2.80   & 3.52   & 2.84  & 2.85  \\
Expt. & 2.87   & 2.82     & 3.52    & -    & -   \\
Other & 2.83   & 2.80   & 3.52   & -    & -       \\
\hline\hline
\end{tabular}
\label{table1}
\end{center}
\end{table}
The calculated lattice parameters of bcc Fe, Ni, fcc Ni, bcc  Ni$_{1}$Fe$_{26}$, 
and Ni$_{1}$Fe$_{15}$ in the NM and FM are presented in Table \ref{table1}. 
The lattice constant of bcc Fe is calculated to be $2.76$\,{\AA} ($2.83$\,{\AA}) 
in the NM (FM) state which is comparable with the previous calculations of 
$2.84$\,{\AA} in the FM state and 
$2.76$\,{\AA} in the NM state\cite{guo} 
and the experimental observations of $2.87$\,{\AA}.\cite{Kittel}
The equilibrium lattice constant of Ni$_{1}$Fe$_{26}$ is determined
to be $2.76$ ($2.84$)\,{\AA} in the NM (FM) state. 
The corresponding values for Ni$_{1}$Fe$_{15}$ are determined 
to be $2.76$ ($2.85$)\,{\AA} in the NM (FM) state. 
Table~\ref{table1} shows that the lattice constant of  bcc Fe increases with the Ni concentrations.
This is consistent with experimental observations,  
which show that Ni as a solute slightly expands the lattice constant of bcc Fe.\cite{mckeehan}
It is interesting to find that the lattice parameter of bcc Fe is not affected by the Ni addition in the NM states;
this can be attributed to the magnetovolume effect.\cite{kubler}
Therefore,  Ni expands the lattice of bcc Fe in the FM state, 
but not in the NM state which suggests that the ferromagnetic interactions are responsible for 
the increment of the lattice constant of Fe with the Ni concentrations. 
The present  calculated values are slightly smaller than those determined 
at room temperature as might be expected from thermal expansion,\cite{ZWELL}
but it is encouraging that  the trend of the lattice parameter as a function of 
the nickel concentration is correctly reproduced.\cite{ZWELL,REED}
It is also noticeable that \textit{ab initio} calculations usually
give trends and it may underestimate or overestimate the lattice parameters. 
Our previous calculations on bcc Fe-based materials show that 
atomic relaxation around the impurities is negligible.\cite{RahmanPRB2010,RahmanFeAl}

\begin{figure}[ht]
\includegraphics[width=10cm]{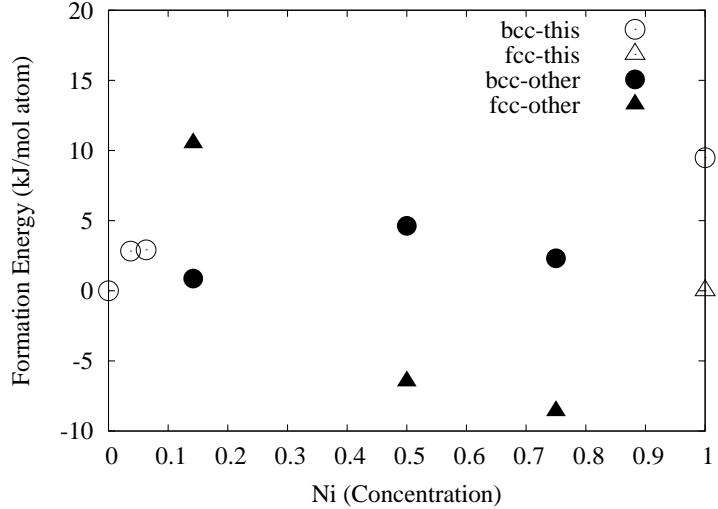}
\caption{Calculated formation energy (kJ/mol atom) of 
bcc Fe, Ni,  Ni$_{1}$Fe$_{15}$, Ni$_{1}$Fe$_{26}$, and fcc Ni versus the Ni concentration. 
Open symbols represent our work (this) and filled symbols show other calculated values (other) 
taken from Ref.~\onlinecite{mishin}. 
Circles (triangles) indicate the bcc (fcc) systems.} 
\label{formation}
\end{figure}

Using the optimized lattice parameters, $\Delta H$ was calculated using Eq. (\ref{formationenthalpy}) and  
the results are given in Fig.~\ref{formation}, which shows $\Delta H$ of fcc NiFe systems 
taken from Ref.~\onlinecite{mishin}.
Note that $\Delta H$ of a system is nothing more than the total energy of the system 
at zero pressure and zero kelvin at the corresponding equilibrium lattice parameter. 
The formation enthalpy is measured relative to bcc Fe and fcc Ni. 
We see that the small addition of Ni in bcc Fe increases $\Delta H$. 
Based on previous~\cite{mishin} and present study,it can be summarized that 
the addition of Ni to Fe stabilizes fcc structures and destabilize the bcc structures, 
consistent with the experimental phase diagram.\cite{owen}

\subsection{Elastic Properties and Thermodynamics}
\begin{table}
\caption{The calculated elastic constants and bulk moduli 
in the units of {GPa} of bcc Fe, fcc Ni, bcc Ni, and bcc Ni$_{1}$Fe$_{26}$.
The experimental (Expt.) values for the bcc Fe \cite{Kittel} and fcc Ni \cite{Kittel} 
are given for comparison.}
\begin{center} 
\begin{tabular}{cccccccccccc}
\hline \hline
& $C_{11}$ & $C_{12}$ & $C_{44}$ &  $G_{\left\{100\right\}}$ &$B$ \\
\hline
bcc Fe &  $\;\;\;257$  & $\;\;\;134$  & $\;\;\;95$ & $\;\;\;61.00$ & $\;\;\;175.00$ \\
Expt.&    $\;\;\;242$  & $\;\;\;136$  & $\;\;\;112$ & $\;\;\;53.00$ & $\;\;\;171.33$ \\ \hline
fcc Ni &  $\;\;\;276$  & $\;\;\;156$  & $\;\;\;130$ & $\;\;\;60.00$ &
$ \;\;\;196.00$ \\
Expt.&   $\;\;\; 248$  & $\;\;\;155$  & $\;\;\;124$ & $\;\;\;46.50$ & $\;\;\;186.00$ \\ \hline
bcc Ni &  $\;\;\;141$  & $\;\;\;211$  & $\;\;\;154$ & $-35.00$ &
$ \;\;\;187.67$ \\ \hline
Ni$_{1}$Fe$_{26}$ &  $\;\;\;267$  & $\;\;\;149$  & $\;\;\;102$ &
$ \;\;\;59.00$ & $\;\;\;188.33$ \\ \hline
Ni$_{1}$Fe$_{15}$ &  $\;\;\;243$  & $\;\;\;127$  & $\;\;\;110$ &
$ \;\;\;58.00$ & $\;\;\;165.67$ \\
\hline\hline
\end{tabular}
\label{table2}
\end{center}
\end{table} 
In order to obtain the elastic properties,  $\pm 0.03$ strains were imposed 
as discussed in Eqs.~(\ref{cprime}) and (\ref{cprime2}).  
The calculated bulk moduli and elastic constants of bcc Ni$_{x}$Fe$_{1-x}$ and 
fcc Ni are shown in Table \ref{table2}. 
The calculated elastic constants of bcc Fe and fcc Ni are close to the experimental values, and confirm that
the number of \textbf{k}-points and the number of basis functions,
used in these calculations, are sufficient to reproduce the experimental data.\cite{Seo}
All the systems have positive elastic constants 
which satisfy the mechanical stability condition of a crystal.\cite{Wohlfarth}
However, bcc Ni is mechanically unstable due to the negative estimated value of $G_{\left\{100\right\}}$.
Although it is mechanically unstable, bcc Ni has been successfully achieved as 
a thin film on GaAs substrate.\cite{bccNi}
If $G_{\left\{110\right\}}$ is larger than  $G_{\left\{100\right\}}$, 
then it is easier to shear on the $\left\{110\right\}$ plane than 
on the $\left\{100\right\}$ plane.  
{It is noticeable in Table \ref{table2}  
both the bulk modulus and elastic constants($C_{11}$ and $C_{12}$) of bcc N$_{1}$Fe$_{26}$ are increased.} 
With the addition of Ni in bcc Fe, $G_{\left\{110\right\}}$ is increasing 
while $G_{\left\{100\right\}}$ is decreasing.

\begin{table}
\caption{The calculated shear modulus $G$ 
(in units of GPa), Young's modulus $E$ (in units of GPa), $B/G$, 
Cauchy pressure $C_{P}$(in units of GPa), anisotropy factor $A$, 
and  Poisson's ratio $\nu$ of bcc Fe, bcc Ni, fcc Ni, and 
bcc Ni$_{1}$Fe$_{26}$ and Ni$_{1}$Fe$_{15}$.}
\begin{center} 
\begin{tabular}{cccccccccccccccccc}
\hline \hline
& $G$ & $E$ & $B/G$ & $C_{P}$& $A$ & $\nu$&\\ \hline
bcc Fe &  $\;\;\;\;\;79.80$  & $\;\;\;207.82$  & $\;\;\;\;\;2.19$ &$ \;\;\;39$ & $\;\;\;\;\;1.54$& $\;\;\;0.30$ &  \\
bcc Ni &  $\;\;-27.18$  & $\;\;-85.67$  & $\;\;-6.90$ & $\;\;\;57$ & $\;\;\;-4.40$& $\;\;\;0.58$ &  \\
fcc Ni &  $\;\;\;\;\;95.32$  & $\;\;\;246.07$  & $\;\;\;\;\;2.06$ & $\;\;\;26$ & $\;\;\;\;2.17$& $\;\;\;0.29$ &  \\
Ni$_{1}$Fe$_{26}$ &  $\;\;\;\;\;81.89$  & $\;\;\;214.57$  & $\;\;\;\;\;2.30$ &$\;\; 47$ & $\;\;\;\;1.73$& $\;\;\;0.30$ &  \\
Ni$_{1}$Fe$_{15}$ &  $\;\;\;\;\;85.08$  & $\;\;\;217.94$  & $\;\;\;\;\;1.95$ & $\;\;\;17$ & $\;\;\;\;1.90$& $\;\;\;0.28$ &  \\
\hline \hline
\end{tabular}
\label{table3}
\end{center}
\end{table}

Table \ref{table3} lists the shear $G$ and 
Young's modulus $E$, $B/G$, Cauchy pressure $C_{P}$, 
anisotropic factors $A$, and Poisson's ratio $\nu$ for Ni$_{x}$Fe$_{1-x}$. 
Note that $G$ is bounded either by $G_{V}$ or $G_{R}$ and,
therefore, $G$ is in an approximated average sense.
It is a common practice to consider $G$ as 
an indication of mechanical properties of materials.\cite{Hatcher,Wagner} 
It is known that the hardness and strength of materials are related to their elastic moduli, 
such as $E$, $B$, and $G$.\cite{Wohlfarth}
The bulk modulus of bcc Fe is slightly increased by 3.7\,at.\% Ni. 
It is noticeable that both $E$ and $G$  increase with Ni. 
On the other hand the negative values of bcc Ni demonstrate that 
this phase violates the mechanical stability condition.\cite{Wohlfarth}
All these materials have high shear anisotropy factor and 
anisotropic behavior as increasing Ni concentrations.
Pettifor\cite{Pett} has suggested that the Cauchy pressure could be used to describe 
the angular character of atomic bonding in metals and compounds. 
If the bonding is more metallic, the Cauchy pressure will be positive. 
It is believed that ductile materials (such as Ni or Al) have positive values and 
brittle materials (such as Si) have negative values of Cauchy pressure.\cite{Pett}
This can be seen in Table \ref{table3} where Ni$_{x}$Fe$_{1-x}$  have positive values 
and are ductile, consistent with the metallic behavior discussed below. 

To further shed light on whether Ni$_{x}$Fe$_{1-x}$  are brittle or ductile, 
we used simple Pugh relations that link empirically the plastic properties of metals 
with elastic moduli by $B/G$.\cite{Pugh} 
If the $B/G$ ratio is greater than $1.75$, the material behaves in a ductile manner, 
otherwise in a brittle manner, as demonstrated by first-principles calculations.\cite{Pett, Chen}
It is seen that the $B/G$ ratio is greater than $1.75$ for bcc Ni$_{x}$Fe$_{1-x}$ pointing that 
all our considered systems are ductile as expected.
The other factor that measures the stability of a crystal against shear is 
Poisson's ratio $\nu$, and it has been shown\cite{Gsch, Yoo} that 
brittle materials have higher Poisson's ration such as NiAl ($\nu=0.41$), 
while the ductile materials have Poisson's ratio $\nu \sim 0.30$. 
Now it is clear from Table~\ref{table3} that the Poisson's ratios of bcc Ni$_{x}$Fe$_{1-x}$ are 
 $\sim 0.30$. 
Therefore, bcc Ni$_{x}$Fe$_{1-x}$ enter into the class of ductile materials. 
At low temperature the vibrational excitations arise solely from acoustic vibrations 
and the calculate isotropic mean velocities, using Eq.~(\ref{vmeq}), are given in Table~\ref{table4}. 

\begin{table}
\caption{The calculated sound velocities (in units of km/s) and 
densities (in units of kg/m$^{3}$) of 
bcc Fe, fcc Ni, and  bcc Ni$_{1}$Fe$_{26}$ and Ni$_{1}$Fe$_{15}$.}
\begin{center} 
\begin{tabular}{cccccccccccccccccc}
\hline \hline
& $v_{\bot}$ & $v_{\|}$ & $v_\mathrm{m}$ & $\rho$&  & \\ \hline
bcc Fe & $\;\;\;5.86$ & $\;\;\;3.12$ & $\;\;\;3.48$ & $\;\;\;8190$ &  \\
fcc Ni & $\;\;\;6.01$ & $\;\;\;3.27$ & $\;\;\;3.64$ & $\;\;\;8940$ & \\
Ni$_{1}$Fe$_{26}$ & $\;\;\;6.05$ & $\;\;\;3.18$ & $\;\;\;3.55$ & $\;\;\;8120$ &  \\
Ni$_{1}$Fe$_{15}$ & $\;\;\;5.89$ & $\;\;\;3.25$ & $\;\;\;3.62$ & $\;\;\;8040$ &  \\
\hline \hline
\end{tabular}
\label{table4}
\end{center}
\end{table}
{We emphasise that the modulus ratio is just one of many indicators of
the propensity of a material to be brittle;\cite{Cottrell}
crystal defects, the weakening of grain boundaries by chemical segregation
and many other factors can play a seminal role in determining the macroscopic properties.
We found that the calculated elastic constants are in agreement with the available data, and 
the knowledge of elastic constants 
permits the sound velocities to be estimated as a function of the direction. 
Simply using Eq.~(\ref{sound1}--\ref{sound3}), 
the sound velocities of all the systems considered here were calculated, 
and the velocities of bcc Ni$_{x}$Fe$_{1-x}$ in the [100], [110], and [111] directions are given 
in Table~\ref{table5}. 

\begingroup
\begin{table}
\caption{The calculated sound velocities (in units of km/s) of 
bcc Fe, fcc Ni, and  bcc Ni$_{1}$Fe$_{26}$ and Ni$_{1}$Fe$_{15}$ 
in different directions. 
The values of $v_\mathrm{m}$ are calculated by using Eq.~(\ref{vmeq}).}
\begin{tabular}{ccccccccccccccc}
\hline\hline \multicolumn{4}{c}{$v_\mathrm{l}$} &&& 
\multicolumn{3}{c}{$v_\mathrm{t}$} &&& 
\multicolumn{3}{c}{$v_\mathrm{m}$}\\
\hline
& [100]\; & [110]\; & [111]\; &&& [100]\;& [110]\; & [111]\;&&& [100]\; & [110]\; & [111]\; \\
\hline
bcc Fe & $5.60$&	$5.96$	&$6.07$&&&		$3.41$&	$2.74$	&$2.98$&&&		$4.34$&	$3.72$&	$4.00$\\
fcc Ni & $5.56$&	$6.22$ &	$6.43$ &&&		$3.81$ &	$2.59$ &	$3.05$ &&&		$4.66$ &	$3.57$&	$4.13$\\
Ni$_{1}$Fe$_{26}$ & $5.73$	&$6.18$&	$6.32$	&&&	$3.54$	&$2.70$	& $3.01$	&&&	$4.49$&	$3.69$&	$4.06$\\
Ni$_{1}$Fe$_{15}$& $5.50$&	$6.06$&	$6.23$&&	&	$3.70$	&$ 2.69$	& $3.06$ &&&	$4.55$	& $3.67$&	$4.11$\\
\hline\hline
\end{tabular}
\label{table5}
\end{table}
\endgroup
One can see that $v_\mathrm{m}$ in the [100] direction, 
which is also an easy axis of magnetization of bcc Fe, is increasing with the Ni concentrations. 
On the other hand,
Ni concentrations decrease $v_\mathrm{m}$ in the [110] direction. 
It is clear to see that $v_\mathrm{m}$ has larger values in the [100] direction than in the [110] direction. 
The different values of $v_\mathrm{m}$ in the different crystallographic directions 
confirm the anisotropic behavior of bcc Ni$_{x}$Fe$_{1-x}$.

Before going to discuss the Bain path, it is emphasized that the elastic constants calculations did not
allow internal coordinate or stress-tensor optimization at each lattice distortion. 
Since lattice distortions were small $\sim \pm 3\%$, this constraint may
preserve the stable states and the effect of atomic relaxation (in either direction, \textit{i.e.}, $x,y,z$) 
is found to very small. 
However, we also performed some test calculations on the internal coordinate relaxations in 
the distorted systems, but they were hardly distinguishable with the unrelaxed ones.
It is  believed that atomic relaxation will not have much effect (qualitatively) 
on the elastic properties and such relaxation will not change 
the main conclusions drawn in this paper.

\subsection{Magnetism on the Bain deformation}
\begin{figure}
\includegraphics[width=5cm]{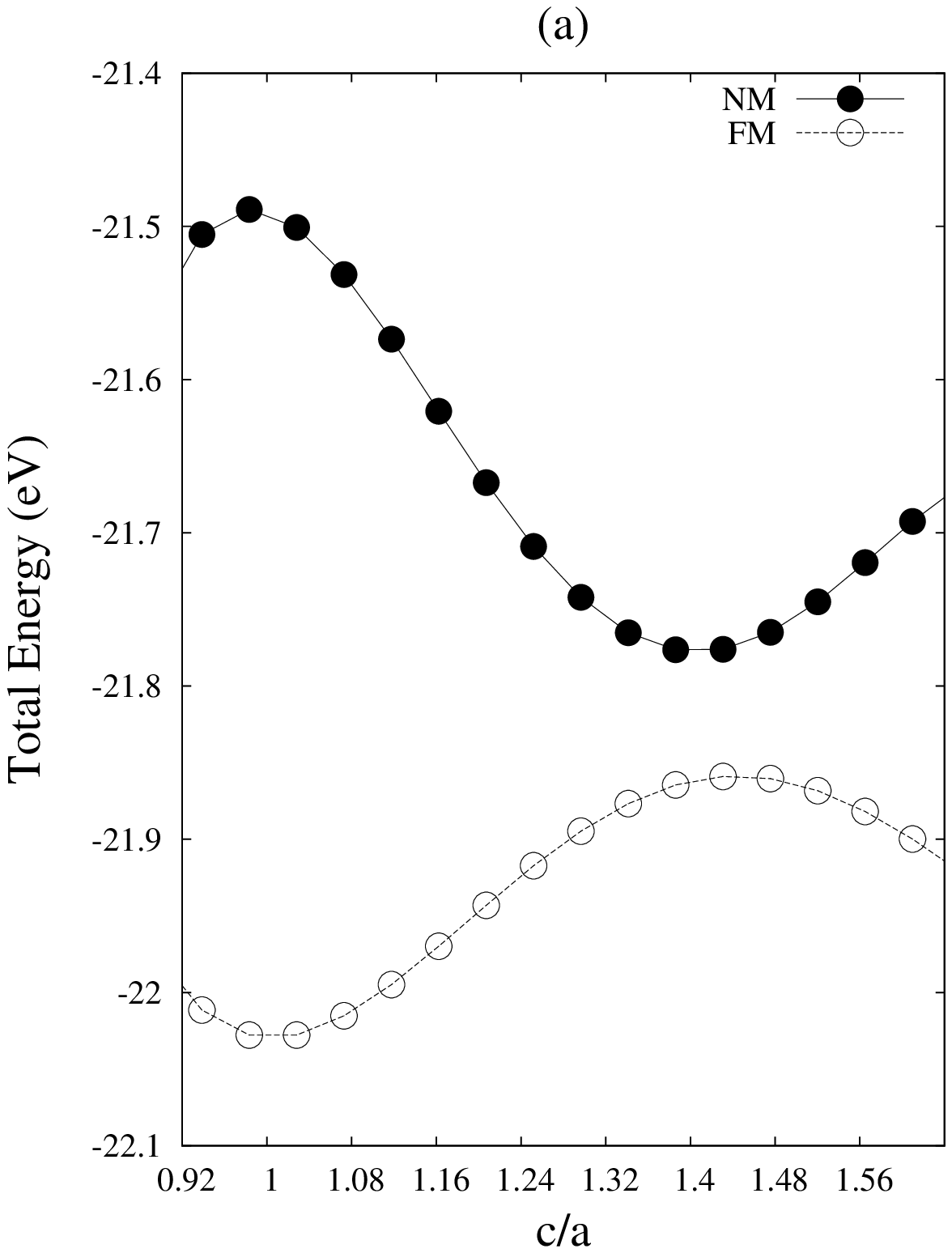}
\includegraphics[width=5cm]{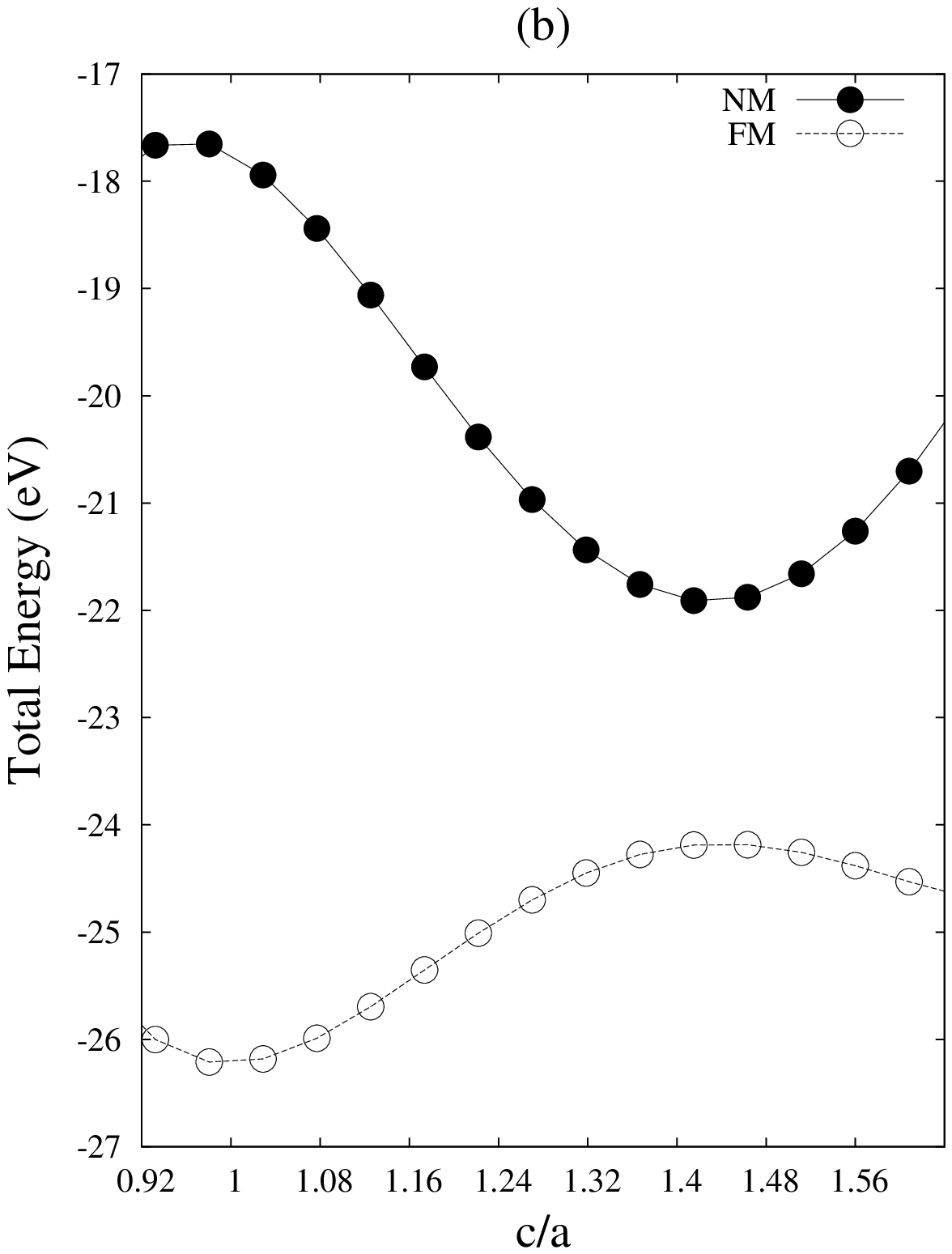}
\includegraphics[width=5cm]{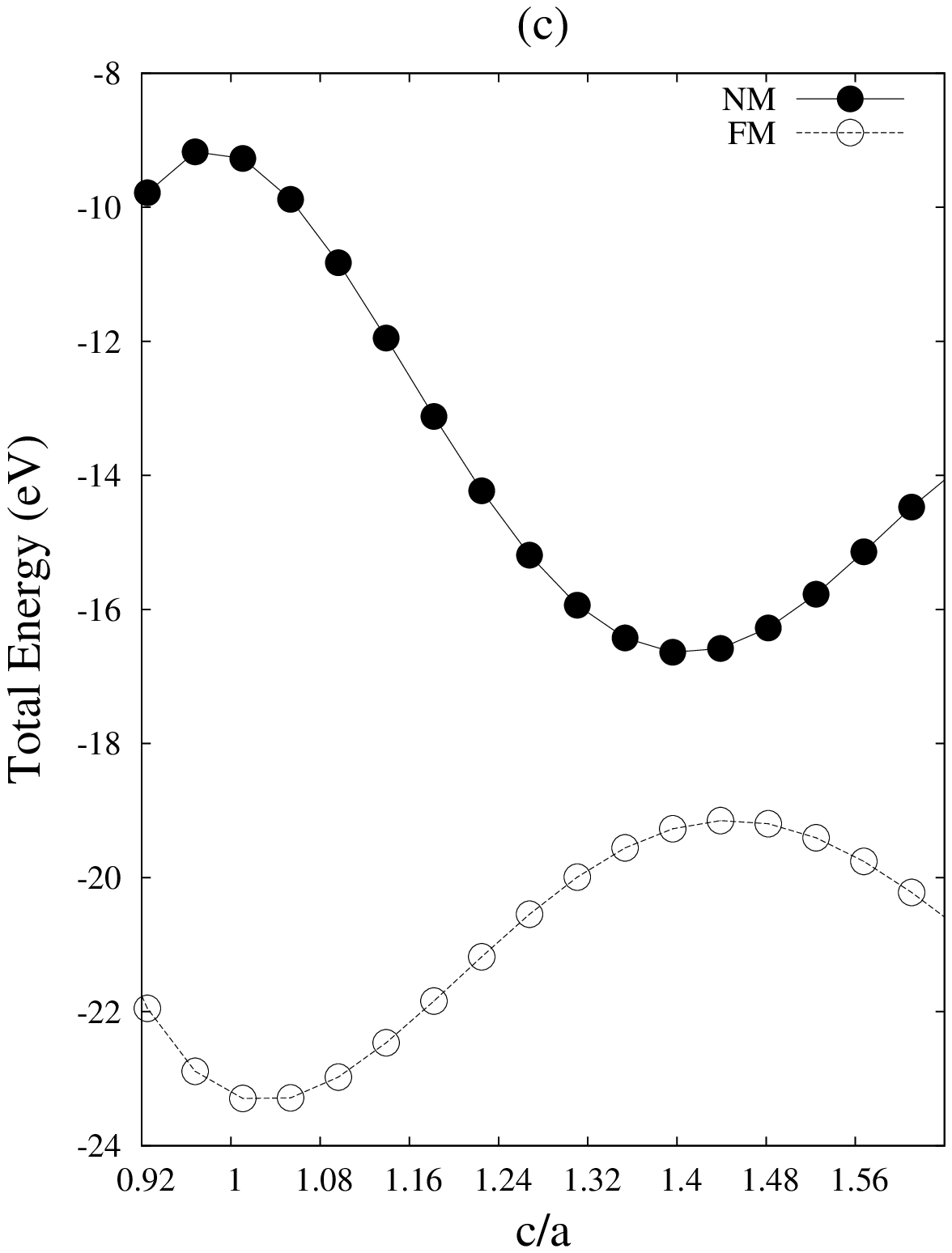}
\caption{Total energy (in eV units) as a function of $c/a$ for 
(a) bcc Fe (b) Ni$_{1}$Fe$_{26}$, and (c) Ni$_{1}$Fe$_{15}$. 
Filled (Open) circles  represent the total energy in NM (FM) states. }
\label{strainfig}
\end{figure}
Let us examine the effects of magnetism
on the martensitic transformation in terms of the Bain deformation.\cite{Krasko}
One should not be confused with the reversibility of martensitic transformation by crystal symmetry,\cite{nature}
because we ignored the effects of entropy totally, which causes the irreversible processes.\cite{Christian}
The mechanical stability of the cubic phase under tetragonal distortion was evaluated 
by calculating the total energy as a function of volume conserving tetragonal strain 
at the optimized lattice parameters of bcc Ni$_{x}$Fe$_{1-x}$.  
The total energies in the NM and FM states as a function of $c/a$ ratio 
at constant volume are presented in Fig.~\ref{strainfig}. 
It is seen that in the NM states the total energy has 
a negative curvature around $c/a=1.0$, 
which is a  signature of mechanical instability.\cite{Wohlfarth}
It is further expected that the structure instability will be accompanied by 
a softening of phonons.\cite{Hsueh,Kittel,Gruner,Herper}
These figures clearly show that when the cubic symmetry of the bcc phase is broken 
by the tetragonal distortion in the NM state, 
the total energy is decreasing and indicating a phase transition 
to the other phase; fcc in this case. 
The total energy is minimized around $c/a=\sqrt{2}$ (fcc phase) and 
become mechanically stable. 
This trend can be seen in all these bcc FeNi systems. 

However, the situation was drastically changed when we repeated 
the same calculations in the FM states. 
In the FM states, these systems are mechanically stable 
in the bcc  phases and unstable in the fcc phases. 
It is possible on the basis of these observations to conclude that 
the bcc phases are stable in the FM states, 
and fcc phases  in the NM states. 
The bcc-fcc phase transition is prohibited by the magnetic contribution to the total energy. 
If these systems were not magnetic, 
then one could expect bcc-fcc phase transitions in these bcc FeNi systems. 
This further points out the role of magnetism in the phase stabilities of FeNi systems, \textit{i.e.}, 
the bcc phases are stabilized by ferromagnetism. 
These findings show that  magnetism contributes to both structural and elastic stabilities of materials. 
Nevertheless, magnetism can either stabilize an elastically unstable material,
as discussed above, or cause a stable structure to become elastically unstable,
\textit{e.g.}, hcp Fe. 
A similar Bain transformation path was also observed for bcc-fcc Fe 
where it has been shown that the local minima on the Bain path was changed by 
changing the lattice volume.\cite{Friak,Okatov}
Therefore, the Bain path can  be altered slightly by changing the lattice volume.

It is worth to mention how our magnetic model describes the reality of alloy system.
Our comparison of the ferromagnetic and nonmagnetic solutions is simply to emphasise
the stability of the former state.
The comparison should ideally be between the ferromagnetic and paramagnetic states
which we cannot easily access.
Alloys differ from the corresponding chemical compounds 
by introducing the concept of atomic thermal aggitation,\cite{Bragg-Williams}
which can be interpreted as an Ising ferromagnet.\cite{Ising}
Of a given atomic configuration at a finite temperature,
the Ni impurity modifies the the magnetic moments of neighboring Fe atoms.
In addition, Fe-Ni system is a representative itinerant ferromagnet\cite{kubler}
described by the Stoner model.\cite{Stoner}
A natural long-range oscillations, so-called the Ruderman-Kittel-Kasuya-Yosida (RKKY)\cite{RKKY}
and Friedel\cite{Friedel} oscillations, may occur.
However, the thermal aggitation will change the atomic configurations dynamically
and it will diminish the effects of the RKKY and Friedel oscillations,
by superposition of the random phases of those oscillations.
Hence, we can expect that the averaged magnetic moments of atoms will not much differ from 
those of a representative atomic configuration.
Although there are good methods for simulating the magnetic moments of disordered states,\cite{DLM}
the problem of alloy magnetism is still open.
In addition, bcc iron matrix screens well such oscillations
of the magnetic moments within the nearest neighbor distance.\cite{RahmanPRB2010,Rahman2008,RahmanFeAl}
Therefore, our model should not change the essence of the conclusions,
since the ferromagnetic state of iron and its dilute alloys is well-established
to be so much more stable than the paramagnetic state.

\subsection{Electronic structures on the Bain deformation}

The observed stability/instability in fcc/bcc phase ultimately have
 an underlying electronic origin. 
The electronic structures of these bcc systems and 
the calculated total electronic density of states (DOS) per atom at the Fermi level ($E_\mathrm{F}$),
denoted as $n\left( E_\mathrm{F}\right)$, in the NM state, are given in Fig.~\ref{totaldos}. 
The bcc phase, Ni$_{x}$Fe$_{1-x}$ has a minimum $n\left( E_\mathrm{F}\right)$, 
and as it is deformed to body-centered tetragonal (bct) 
then $n\left( E_\mathrm{F}\right)$ suddenly increases and 
has a maximum value around $c/a=1.08$. 
The lowest $n$($E_\mathrm{F}$) can be seen near $c/a=\sqrt{2}$. 
This shows that the bcc structure is mechanically unstable due to large 
$n\left( E_\mathrm{F}\right)$. 
\begin{figure}[]
\includegraphics[width=7cm]{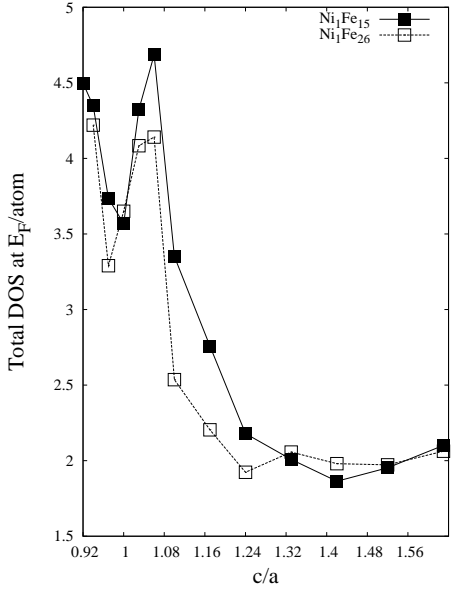}
\caption{Total density of states (DOS) per atom at $E_\mathrm{F}$ 
of bcc  Ni$_{1}$Fe$_{15}$ and Ni$_{1}$Fe$_{26}$ as a function of $c/a$. 
Filled (Open) squares represent the DOS of Ni$_{1}$Fe$_{15}$ (Ni$_{1}$Fe$_{26}$) 
in the NM states.}
\label{totaldos}
\end{figure}

\begin{figure}[]
\includegraphics[width=5cm]{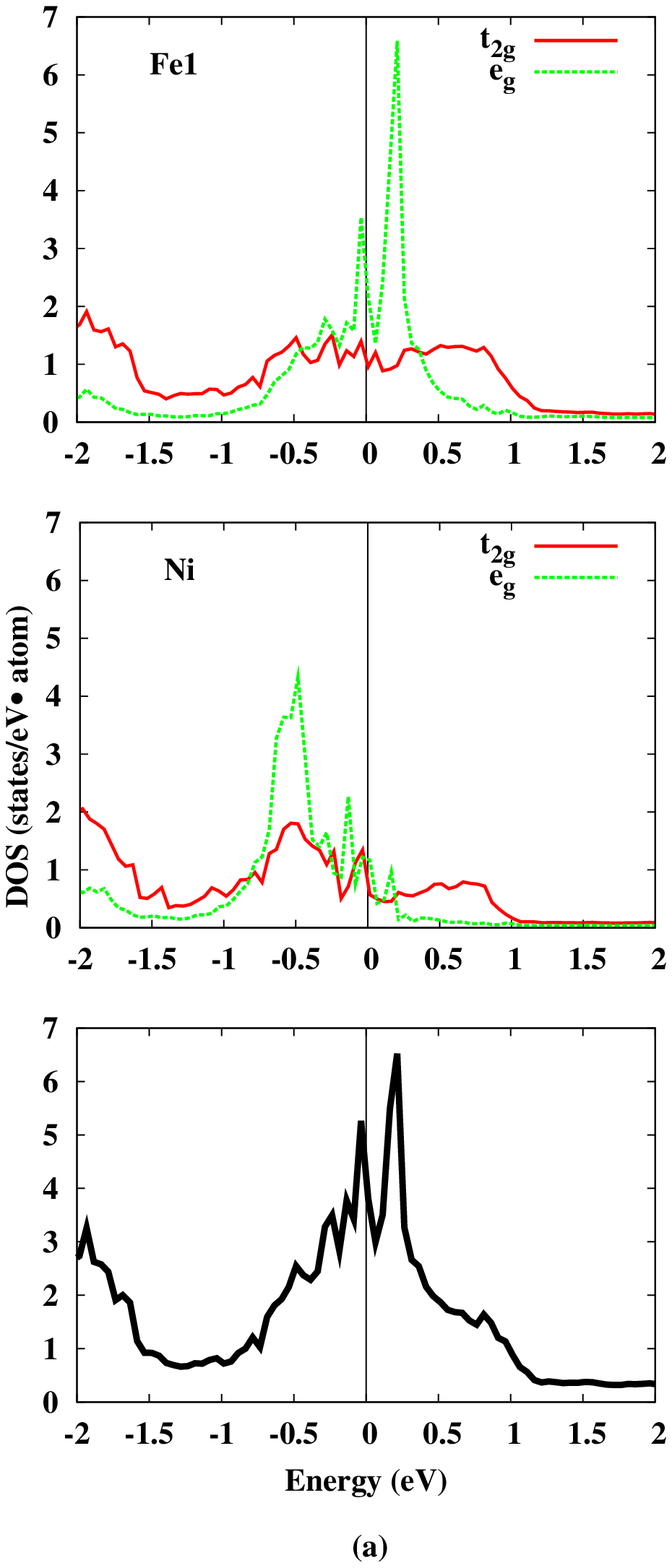}
\includegraphics[width=5cm]{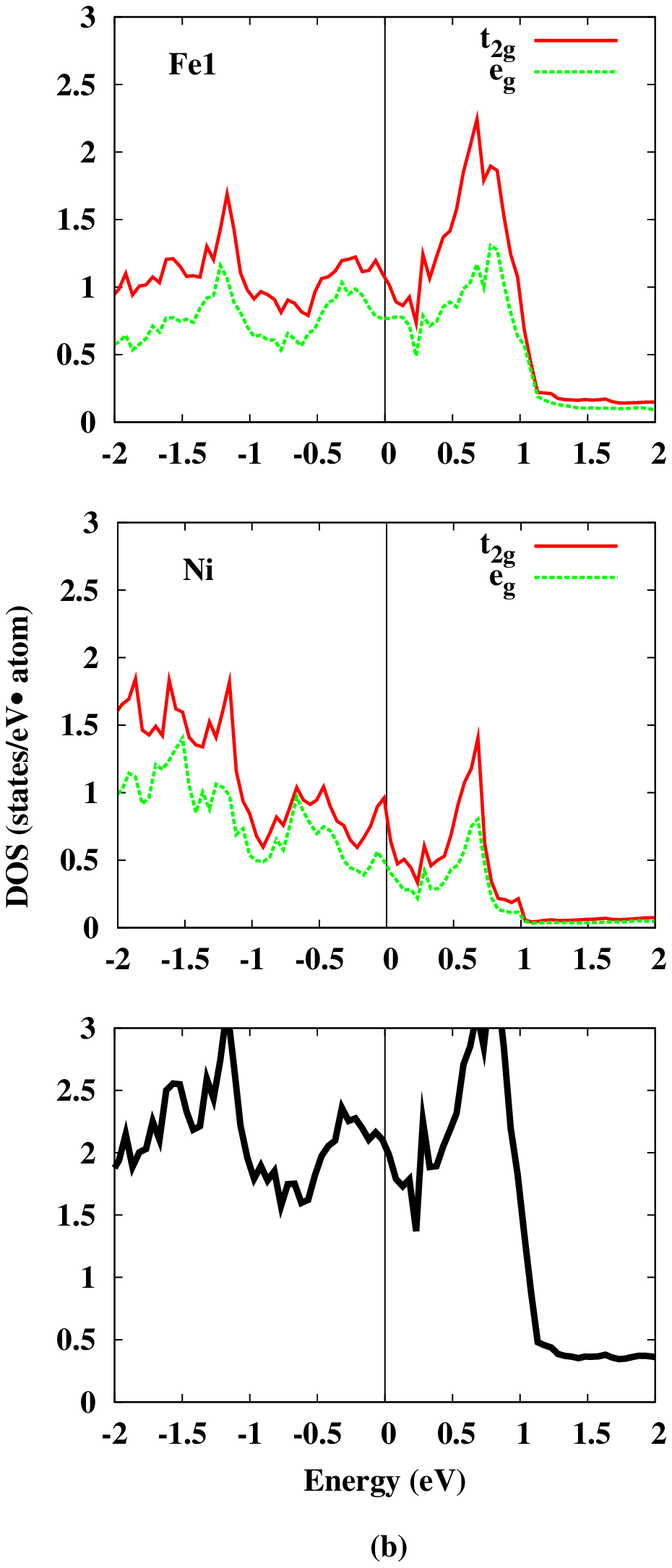}
\caption{(Color online) Calculated spin-unpolarized impurity-atom-projected 
local density of states of Ni$_{1}$Fe$_{26}$ for (a) $c/a=1.0$ 
and (b) $c/a=\sqrt{2}$. 
Solid (dotted) lines represent the $t_{2g}$ ($e_{g}$) states, 
whereas the bold solid lines show the total DOS per atom. 
The Fermi energy ($E_\mathrm{F}$) is set to zero.}
\label{Fe26NiDOS}
\end{figure}

To reveal the effect of tetragonal distortion on the local DOS, 
the atomic projected local DOS of Ni$_{1}$Fe$_{26}$ 
and Ni$_{1}$Fe$_{15}$ is shown in Fig.~\ref{Fe26NiDOS} 
and Fig.~\ref{Fe15NiDOS}, respectively, which show the local DOS for $c/a=1.0$ and $c/a=\sqrt{2}$. 
The DOS shows metallic behavior which also follows the Pettifor\cite{Pett} 
suggestions about the metallic character and positive values of the Cauchy pressure.  
One can clearly see the reduction ($\sim50  \%$) of $n\left( E_\mathrm{F}\right)$
 for the $c/a=\sqrt{2}$ cases. 
In both Ni$_{1}$Fe$_{26}$ and Ni$_{1}$Fe$_{15}$, 
$n\left( E_\mathrm{F}\right)$ is dominated 
by the $e_{g}$ orbitals for $c/a=1.0$, however, 
the $t_{2g}$ orbitals contribute to $n\left( E_\mathrm{F}\right)$ for $c/a=\sqrt{2}$. 
It is also clear from these figures that $E_\mathrm{F}$ of cubic Ni$_{x}$Fe$_{1-x}$ is 
located on a peak of high DOS and the tetragonal distortion is leading 
to a decrease of $n\left( E_\mathrm{F}\right)$ in the tetragonal phase. 
The strong reduction in $n\left( E_\mathrm{F}\right)$ is 
an indicative of a states shifts from the Fermi surface, \textit{i.e.}, 
an instability associated with the peak in $n\left( E_\mathrm{F}\right)$. 
Note that in the absence and presence of distortion, 
the value of $n\left( E_\mathrm{F}\right)$ for bcc Ni$_{x}$Fe$_{1-x}$ is 
within the Stoner limit.\cite{RahmanPRB2010}
In metals, in general, a lower $n\left( E_\mathrm{F}\right)$
corresponds to lower electronic kinetic energy. 
This reduction gives a substantial negative contribution to the electronic energy 
and consequently the total energy decreases upon lattice distortion 
as shown above, yielding a negative curvature of the shear modulus for bcc FeNi systems. 
This mechanism is responsible for the elastic instability of bcc Ni$_{x}$Fe$_{1-x}$ in the NM states. 
\begin{figure}[]
\includegraphics[width=5cm]{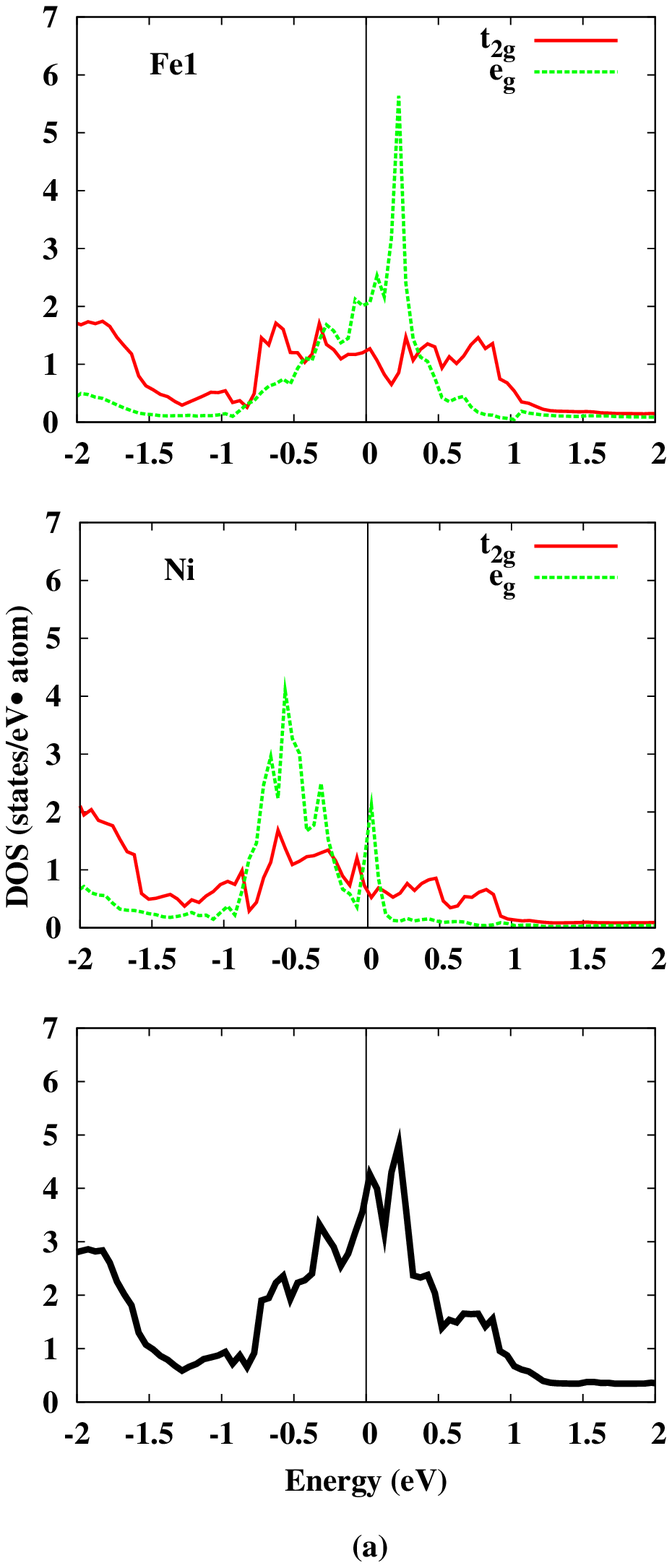}
\includegraphics[width=5cm]{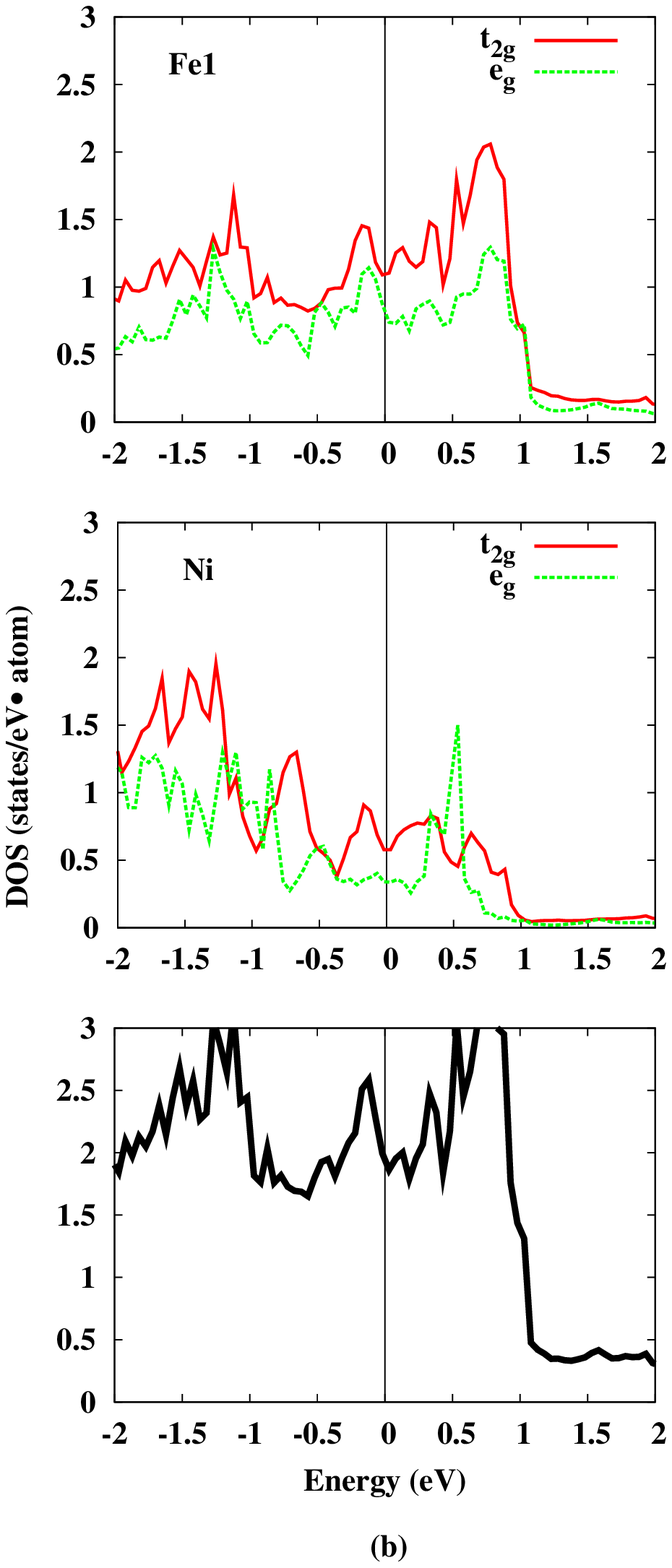}
\caption{(Color online) Calculated spin-unpolarized impurity-atom-projected
local density of states of Ni$_{1}$Fe$_{15}$ for 
(a) $c/a=1.0$ and (b) $c/a=\sqrt{2}$.
Solid (dotted) lines represent the $t_{2g}$ ($e_{g}$) states, 
whereas the bold solid lines show the total DOS per atom. 
The Fermi energy ($E_\mathrm{F}$) is set to zero.}
\label{Fe15NiDOS}
\end{figure}
 
\section{Summary}
The elastic and thermodynamics of Ni impurities in bcc Fe is studied 
using first-principles calculations which showed that Ni impurities do not expand 
the lattice constant of bcc Fe in the nonmagnetic state, 
in contrast to the the ferromagnetic state. 
Nickel impurities improved the elastic properties of bcc Fe. 
The sound velocities of elastic waves were shown to be increasing in the [100] directions, 
and decreasing in the [110] directions with the Ni concentrations. 
We also investigated the Bain path of bcc Ni$_{x}$Fe$_{1-x}$, 
and it was observed that Ni$_{x}$Fe$_{1-x}$ systems are elastically unstable 
for $c/a=1.0$ in the nonmagnetic sates, 
whereas Ni$_{x}$Fe$_{1-x}$ systems are stable for for $c/a=\sqrt{2}$. 
The elastic stability was explained in terms of electronic structures of Ni$_{x}$Fe$_{1-x}$. 

\begin{acknowledgments}
This work was supported by the Steel Innovation Program by POSCO, the WCU (World Class University) program
(Project No. R32-2008-000-10147-0), the Basic Science Research Program (Grant No. 2009-0088216) through 
the National Research Foundation funded by
the Ministry of Education, Science and Technology of Republic of Korea. 
\end{acknowledgments}


\begin{thebibliography}{99}
%
\bibitem{owen} 	E. A. Owen and Y. H Liu, J. Iron St. Inst. \textbf{163}, 132 (1949).
%
\bibitem{williams} A. R. Williams, V. L. Moruzzi, and C. D. Gelatt, J. Magn. Magn. Mater.  \textbf{31}, 88 (1983).
%
\bibitem{meyer}	 R. Meyer and P. Entel, Phys. Rev. B \textbf{57}, 5140 (1998).
%
\bibitem{Olson}  G. L. Krasko and G. B. Olson, J. Appl. Phys. \textbf{67}, 4570 (1990). 
%
\bibitem{Ekman} M. Ekman, B. Sadigh, K. Einarsdotter, and P. Blaha, Phys. Rev. B \textbf{58}, 5296 (1998).
%
\bibitem{Friak} M. M. Fri\'{a}k, M. \~{S}ob, and V. Vitek, Phys. Rev. B \textbf{63}, 052405 (2001).
%
\bibitem{Hsueh} M. H. C. Hsueh, J. Crain, G. Y. Guo, H. Y. Chen, C. C. Lee, K. P. Chang, and H. L. Shih, 
Phys. Rev. B \textbf{66}, 052420 (2002).
%
\bibitem{Okatov} S. V. Okatov, A. R. Kuznetsov, Yu. N. Gornostyrev, V. N. Urtsev, and M. I. Katsnelson, 
Phys. Rev. B \textbf{79}, 094111(2009).
%
\bibitem{filho} J. B. Filho and C. A. Kuhnen, 
Braz. J. Phys. \textbf{23}, 288 (1993).
%
\bibitem{mishin} Y. Mishin, M. J. Mehl and  D. A. Papaconstantopoulos, 
Acta Mater. \textbf{53}, 4029 (2005). 
%
\bibitem{RahmanPRB2010} G. Rahman, I. G. Kim, 
H. K. D. H. Bhadeshia, and A. J. Freeman,  Phys. Rev. B \textbf{81}, 184423 (2010).

\bibitem{flapw} E. Wimmer, H. Krakauer, M. Weinert, and A. J. Freeman, 
Phys. Rev. B \textbf{24}, 6864 (1981).
%
\bibitem{flapw1} M. Weinert, E. Wimmer, and A. J. Freeman, 
Phys. Rev. B \textbf{26}, 4571 (1982).
%
\bibitem{GGA} J. P. Perdew, K. Burke, and M. Ernzerhof, 
Phys. Rev. Lett. \textbf{77}, 3865 (1996).
%
\bibitem{jhlee} J.-H. Lee, T. Shishidou, and A. J. Freeman, 
Phys. Rev. B \textbf{66}, 233102 (2002).
%
\bibitem{monk} H. J. Monkhorst and J. D. Pack, 
Phys. Rev. B \textbf{13}, 5188 (1976).
%
\bibitem{XO} M. Weinert, G. Schneider, R. Podloucky, and J. Redinger,
J. Phys.: Condens. Matter \textbf{21}, 084201 (2009).
%
\bibitem{Seo} S. -W. Seo, Y. Y. Song, G. Rahman, I. G. Kim, M. Weinert, and A. J. Freeman, 
J. Magn. \textbf{14}, 137 (2009).
%
\bibitem{birch} F. Birch, Phys. Rev. \textbf{71}, 809 (1947); 
F. D. Murnaghan, Proc. Nat'l. Acad. Sci. U. S. A. \textbf{30}, 244 (1944).
%

\bibitem{Wohlfarth} G. Grimvall, 
\textit{Thermophysical Properties of Metals} (North-Holland, Amsterdam, 1986).
%
\bibitem{voigt} W. Voigt,  
\textit{Lehrbuch der Kristallphysik.} (Teubner, Leipzig, 1928).
%
\bibitem{reuss}  A. Reuss and Z. Angew, Math. Mech. \textbf{9}, 49  (1929).
%
\bibitem{hill} R. Hill, Proc. Phys. Soc. (London) Ser. A \textbf{65}, 349 (1952).
%
\bibitem{zener} C. Zener, 
\textit{Elasticity and Anelasticity of Metals} (University of Chicago Press, Chicago, 1948).
%
%
\bibitem{Debye} J. D. Walecka, 
\textit{Fundamentals of Statistical Mechanics: Manuscript and Notes of Felix Bloch}
(Imperial College Press, London, 2000).
%
\bibitem{Kittel} C. Kittel, 
\textit{Introduction to Solid State Physics} 7th ed. (Wiley, New York, 1996).
%
\bibitem{guo} G. Y. Guo and H. H. Wang, Chin. J. Phys. \textbf{38}, 949 (2000).
%
\bibitem{mckeehan} L. W. McKeehan, Phys. Rev. \textbf{21}, 402 (1923).
%
\bibitem{kubler} J. K\"{u}bler, 
\textit{Theory of Itinerant Electron Magnetism}
(Oxford University Press, Oxford, 2000).
%
\bibitem{Rahman2008} G. Rahman and I. G. Kim, J. Magn. \textbf{13}, 124 (2008). 
%
\bibitem{ZWELL} L. Zwell, D. E. Carnahan, and G. R. Speich, 
Metal. Trans. \textbf{1}, 1007 (1970).
%
\bibitem{REED} R. P. Reed and R. E. Schramm, J. Appl. Phys. \textbf{40}, 3453 (1969).
%
%
\bibitem{RahmanFeAl} G. Rahman and I. G. Kim, J. Magn. \textbf{16}, 1 (2011). 
%
\bibitem{bccNi} C. S. Tian, D. Qian, D. Wu, R. H. He, Y. Z. Wu, W. X. Tang, L. F. Yin,
Y. S. Shi, G. S. Dong, X. F. Jin, X. M. Jiang, F. Q. Liu, H. J. Qian, K. Sun,
L. M. Wang, G. Rossi, Z. Q. Qiu, J. Shi,
Phys. Rev. Lett. \textbf{94}, 137210 (2005).
%
\bibitem{Hatcher} N. Hatcher, O. Yu. Kontsevoi, and A. J. Freeman, 
Phys. Rev. B \textbf{79}, 020202(R) (2010).
%
\bibitem{Wagner} M. F. -X. Wagner and W. Wind, 
Acta Mater. \textbf{56}, 6232 (2008).
%
\bibitem{Pett} D. G. Pettifor, Mater. Sci. Tech. \textbf{8}, 345 (1992).
%
\bibitem{Pugh} S. F. Pugh, Philos. Mag. \textbf{45}, 823 (1992).
%
\bibitem{Chen} K. Chen, L. R. Zhao, J. Rodgress, and J. S. Tse, 
J. Phys. D: Appl. Phys. \textbf{36}, 435 (2003). 
%
\bibitem{Gsch} K. Gschneidner, A. Russell, A. Pecharsky, J. Morris, Z. Zhang, 
T. Lograsso, D. Hsu, C. H. C. Lo, Y. Ye, A. Slager, and D. Kesse, 
Nature Mater. \textbf{2}, 587 (2002). 
%
\bibitem{Yoo} M. H. Yoo, T. Takasugi, S. Hanada, and O. Izumi, 
Mater Trans. JIM \textbf{31}, 435 (1990). 
%

\bibitem{Cottrell} A. H. Cottrell, Mater. Sci. Technol. \textbf{7}, 981 (1991).

 





\bibitem{Krasko} G. L. Krasko and G. B. Olson, Phys. Rev. B \textbf{40}, 11\,536 (1989).

\bibitem{nature} K. Bhattacharya, S. Conti, G. Zanzotto, and J. Zimmer, Nature \textbf{428}, 55 (2004).

\bibitem{Christian} J. W. Christian, \textit{The Theory of Transformations in Metals and Alloys}
(Pergamon, Amsterdam, 2002) Chap. 4.

\bibitem{Gruner} M. E. Gruner, W. A. Adeagbo, A. T. Zayak, A. Hucht, and P. Entel,
Phys. Rev. B \textbf{81}, 064109 (2010).

\bibitem{Herper} H. C. Herper, E. Hoffmann, P. Entel, and W. Weber,
J. Phys. IV \textbf{5}, C8-293 (1995).

\bibitem{Bragg-Williams} W. L. Bragg and E. J. Williams, Proc. R. Soc. Lond. A \textbf{145}, 699 (1934);
\textit{ibid.} \textbf{151}, 540 (1935); E. J. Williams, \textit{ibid.} \textbf{152}, 231 (1935).

\bibitem{Ising} H. A. Bethe, Proc. R. Soc. Lond. A \textbf{150}, 552 (1935);
T. D. Lee and C. N. Yang, Phys. Rev. \textbf{87}, 410 (1952).

\bibitem{Stoner} E. C. Stoner, Rep. Prog. Phys. \textbf{11}, 43 (1947).

\bibitem{RKKY} M. A. Ruderman and C. Kittel, Phys. Rev. \textbf{96}, 99 (1954);
T. Kasuya, Prog. Theor. Phys. \textbf{16}, 45 (1956);
K. Yosida, Phys. Rev. \textbf{106}, 893 (1957).

\bibitem{Friedel} J. Friedel, Nuovo Cimento Suppl. \textbf{7}, 287 (1958).

\bibitem{DLM} J. Staunton, B.L. Gyorffy, A.J. Pindor, G.M. Stocks, H. Winter,
J. Magn. Magn. Mater. \textbf{45}, 15 (1984); T. Gebhardt, D. Music, M. Ekholm, I. A. Abrikosov, L. Vitos,
A. Dick, T. Hickel, J. Neugebauer, and J. M. Schneider, J. Phys.: Condens. Matter, \textbf{23}, 246003(2011).

\end{thebibliography}
\end{document}